\documentclass[runningheads]{llncs}
\usepackage{subfloat}
\usepackage{graphicx}
\usepackage{floatrow}
\usepackage{multirow}
\usepackage{hyperref}
\newfloatcommand{capbtabbox}{table}[][\FBwidth]

\begin{document}
%
\title{Nuclear Instance Segmentation using a Proposal-Free Spatially Aware Deep Learning Framework}
\titlerunning{SpaNet}
%
 \author{Navid Alemi Koohbanani\inst{1,3}\and
Mostafa Jahanifar\inst{2} \and
Ali Gooya\inst{4} \and
 Nasir Rajpoot \inst{1,3}
}

\authorrunning{Koohbanani et al.}
%
 \institute{University of Warwick, Coventry, UK \and
 NRP Co., Tehran, Iran \and
 The Alan Turing Institute, London, UK \and
 University of Leeds, Leeds, UK
 }

\maketitle              
\begin{abstract}
Nuclear segmentation in histology images is a challenging task due to significant variations in the shape and appearance of nuclei. One of the main hurdles in nuclear instance segmentation is overlapping nuclei where a smart algorithm is needed to separate each nucleus. In this paper,  we introduce a proposal-free deep learning based framework to address these challenges. To this end, we propose a spatially-aware network (SpaNet) to capture spatial information in a multi-scale manner. 
A dual-head variation of the SpaNet is first utilized to predict the pixel-wise segmentation and centroid detection maps of nuclei. Based on these outputs, a single-head SpaNet predicts the positional information related to each nucleus instance.
Spectral clustering method is applied on the output of the last SpaNet, which utilizes the nuclear mask and the Gaussian-like detection map for determining the connected components and associated cluster identifiers, respectively. The output of the clustering method is the final nuclear instance segmentation mask. We applied our method on a publicly available multi-organ data set\footnote{\tiny{\url{https://nucleisegmentationbenchmark.weebly.com/}}} and achieved state-of-the-art performance for nuclear segmentation.

\keywords{Computational pathology  \and Instance Segmentation \and Nuclear segmentation \and }
\end{abstract}
\section{Introduction}
Nuclear segmentation is often the first step toward a detailed analysis of histology images. For instance, automatic nuclear pleomorphism scoring and cancer grading heavily rely on morphological appearance and structure of nuclei, which have been verified by a wide range of studies, see for example \cite{lee2017nuclear}. 
The complexity of nuclei shape and appearance, imperfect slide preparation/staining, and scanning artifacts make the automatic instance segmentation of nuclei hardly achievable.
To overcome these challenges and reduce the burden of manual segmentation, several algorithms have been proposed for nuclear segmentation and detection, ranging from simple thresholding techniques to more sophisticated approaches \cite{vu2019methods}.
However, since the emergence of deep learning and its applications in segmentation, most of these methods have been replaced by Convolutions Neural Networks (CNNs) or are only used as a post/pre-processing step in conjunction with CNNs \cite{vu2019methods}.

Previous methods are mainly based on region-proposal networks, like Mask-RCNN \cite{he2017mask} and PA-Net \cite{liu2018path}, or encoder-decoder neural structures particularly U-Net model \cite{ronneberger2015u}.
Since U-Net was not well established for separating close object in complex histology images, various methods have been introduced in the literature  which concentrates on the following 4 aspects:  \romannumeral 1) modifying the network architecture  to extract richer information (like CIA-Net\cite{zhou2019cia}), \romannumeral 2) introducing auxiliary outputs to the network, the auxiliary output can be the nucleus contour or bounding box (like DCAN \cite{chen2017dcan}, BES-Net \cite{oda2018besnet}), \romannumeral 3) some methods proposed CNNs that predicts distance map (or other geometrical mappings) of nuclei instances (like DR-Net \cite{naylor2018segmentation}), and \romannumeral 4) taking into account different combinations of above-mentioned variations to make their deep learning platform more robust for detecting individual objects \cite{vu2019methods}. Despite these advancements, these models lack spatial awareness which can improve instance-wise segmentation of clustered nuclei, especially in advanced stages of the tumor.

Here, we introduce a novel proposal-free deep learning framework (SpaNet) that can predict spatial information of each nucleus. Outputs of SpaNet are then post-processed through a simple yet effective clustering algorithm to achieve instance-level segmentation.
Our contributions can be summarized as:
\romannumeral 1) a deep learning based proposal-free framework for nuclei instance segmentation having low computational cost and simple post-processing steps inspired by \cite{liang2017proposal},
\romannumeral 2) a spatially-aware network architecture, which is equipped with a novel multi-scale dense convolutional unit, 
\romannumeral 3) incorporating a nuclei detection map for estimating the number of clusters per nuclei clump, 
\romannumeral 4) achieving state-of-the-art results on a well-known publicly available multi-organ data set.

Details for the methodology of the above-mentioned contributions are described in section \ref{Sec:Method}, experimental setups, results and discussion are elaborated in section \ref{Sec:Results}, and finally paper is concluded in section \ref{Sec:Conclusion}.
\vspace{-1mm}

\section{Methods}
\vspace{-1.5mm}
\label{Sec:Method}
Our proposed method consists of predicting positional information of each nucleus through a spatial aware CNN, and then clustering that information to construct instance-level segmentation. To achieve a reasonable spatial prediction and to estimate the number of clusters in nuclei clumps, we additionally incorporated a dual-head network for nuclei mask segmentation (semantic level) and detection maps. In this section, we firstly describe the network architecture, which is used throughout our framework. Afterward, details of employing the proposed CNN for instance segmentation will be discussed.

\subsection{Spatially Aware Neural Network}
An essential step in our proposed nuclear instance segmentation framework is predicting the positional information of each nuclei using CNNs. Conventional CNNs cannot capture positional details due to the nature of kernels.  Convolutional kernels in common CNN architectures extract local features. Hence they give no intuition about the relative position of objects (detected features) in the image.
To address this issue, we propose spatial information aware CNN capable of capturing positional information in all layers. By providing the network with positional information ($x$ and $y$ image coordinates) in the input and keeping that information available to all convolutional kernels, spatial awareness is guaranteed. Details about the positional information in the input and structuring element of the network are discussed in the following sections. 

\subsubsection{Structuring blocks}

\begin{figure}[t]
\includegraphics[width=1.0\textwidth]{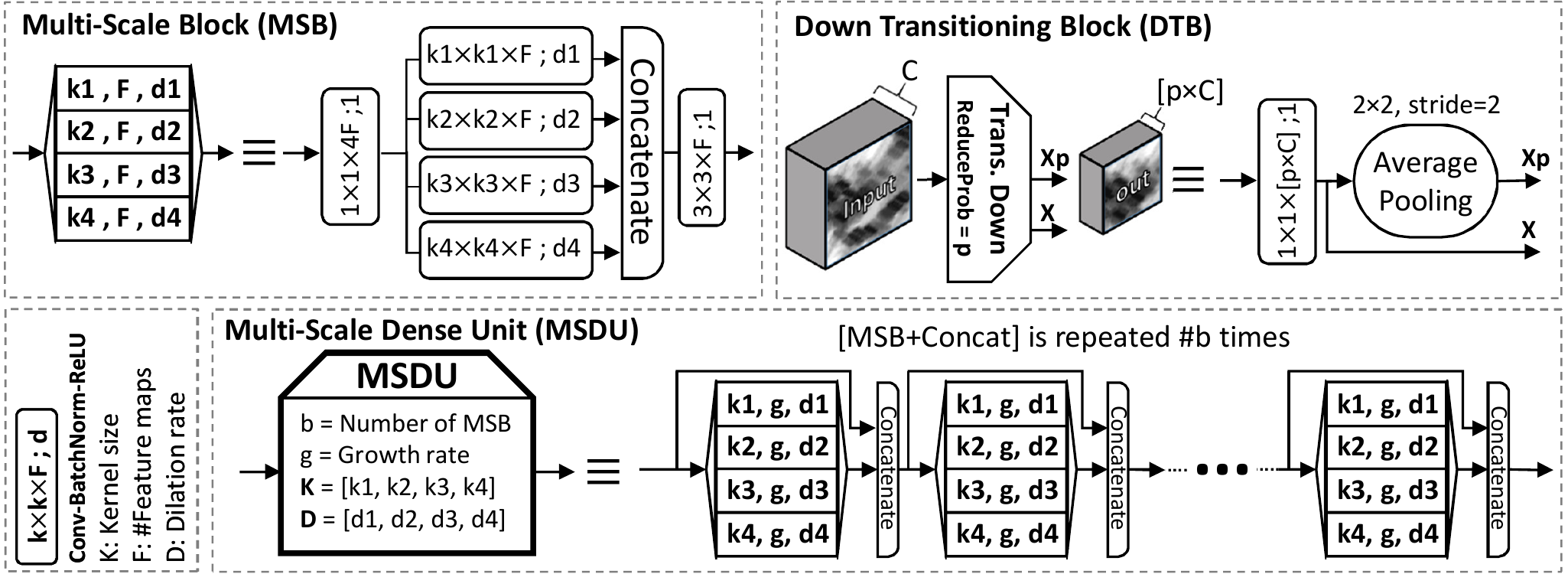}
	\caption{Structuring blocks used in the SpaNet architecture.
	}
\label{fig:BuildingBlocks}
\end{figure}
Preserving spatial information throughout the network is feasible using our proposed multi-scale dense unit (MSDU). MSDU is a densely connected building block inspired by \cite{huang2017densely}. Unlike the ordinary dense unit, our proposed MSDU benefits from the multi-scale convolutional block (MSB) \cite{jahanifar2018segmentation}. Fig. \ref{fig:BuildingBlocks} demonstrates the configuration of a single MSB composed of four parallel convolutional blocks (convolution layer followed by batch-normalization and ReLU layers) with varying kernel size. Having the flexibility to stack convolutional blocks with varying kernel (dilation) rates allows us to obtain multi-resolution feature maps, leading to better performance. MSB Blocks are configured with a specific number of channels ($F$), kernel sizes ($\bf{k}$), and dilation rates ($\bf{d}$). Each MSB block has a $1\times1$  and  a $3\times3$ convolutional block in its terminals to reduce the number of processed and generated feature maps.

As depicted in Fig. \ref{fig:BuildingBlocks}, concatenation layers in the MSDU aggregate the output feature maps from their preceding MSBs. 
The feature aggregation property of MSDUs enables the proposed instance detection network in Fig. \ref{fig:SpaNet} to preserve the positional information (which were passed to the network’s input) at all convolutional blocks throughout its path, making it a spatial aware network.
An MSDU has four configuring parameters: growth rate ($g$), which indicates the number of feature maps generated by every MSB inside the MSDU. $\bf{K}$ and $\bf{D}$ vectors showing the kernels’ sizes and dilation rates of MSB blocks, and $b$ denotes the number of MSB and concatenation pair repetitions. It has been shown that restricting the number of extracted features in each convolutional blocks (setting small growth rates) and aggregating the feature maps instead, result in better performance while reducing the computational costs \cite{huang2017densely}.
 
Other two structuring blocks are Down Transitioning Block (DTB) and  Up Transitioning Block (UTB) which down-sample and up-sample their input feature by the scale of 2, respectively. The structure of a DTB is shown in Fig. \ref{fig:BuildingBlocks}, which comprises a $1\times1$ convolutional block that generates $\left[{p \times C}\right]$ feature maps ($\bf{X}$). The parameter $C$ is the number of input feature maps to the DTB, and $0<p<1$ is the reducing rate. DTB also consists of a $2\times2$ average pooling layer with a stride of 2, which will down-sample the size of feature maps in half ($\bf{Xp}$). UTB comprises a $2 \times 2 \times {\left[{p \times C}\right]}$ transposed convolution layer followed by batch-normalization and ReLU layers.

\subsubsection{Spa-Net architecture}
\begin{figure}[t]
\includegraphics[width=1.0\textwidth]{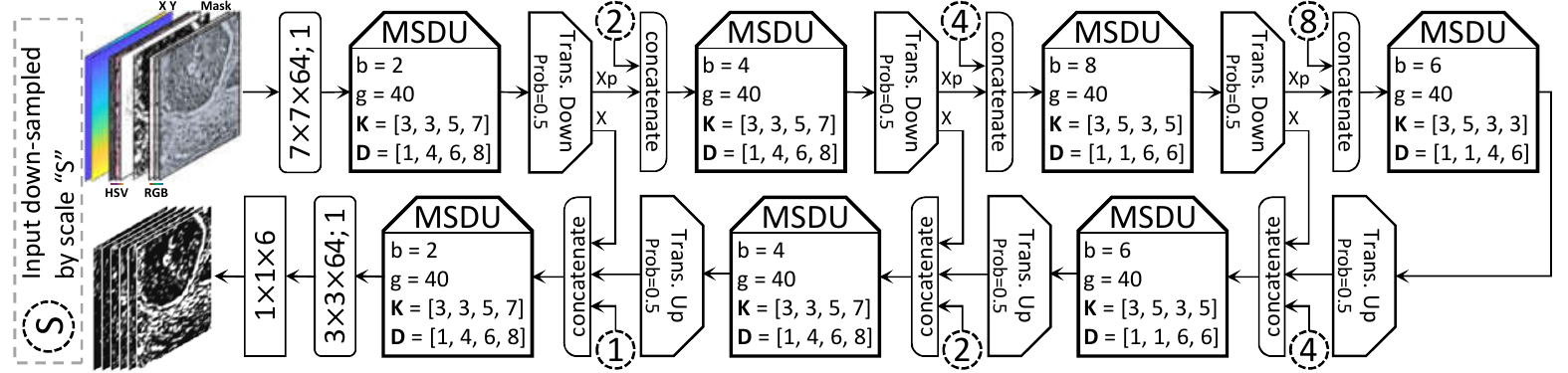}
	\caption{Overview of the SpaNet architecture.}
\label{fig:SpaNet}
\end{figure}
The proposed spatial aware network for nuclei instance segmentation, SpaNet, is illustrated in Fig. \ref{fig:SpaNet}. The main structure in SpaNet is MSDU, which is equipped with a feature aggregation property that enables positional information flows throughout the network.
 Feature maps in SpaNet are down-sampled three times by DTBs in the encoding path and are up-sampled accordingly by UTBs in the decoding path. Skip connections will make the feature maps in the decoding path more spatially enriched and facilitate gradient flow during training \cite{ronneberger2015u}. More importantly, there are some points in the network that we lose direct access to the positional information (after DTB and UTB units) where feature aggregation is not applied. As a workaround, we appropriately scaled the network input and added it in these layers via concatenation layers.

As shown in Fig. \ref{fig:SpaNet}, configuring parameters of each MSDU is different, except for the growth rate ($g$).
Other parameters are tuned based on the MSDU position in the network. An advantage of the SpaNet is capturing small-to-large structures in all levels by appropriately setting the MSDUs' parameters. At the first level of the SpaNet where feature maps and nuclei regions are relatively large, MSDU kernel sizes and dilation rates are set to ${\bf{K}}=[3, 3, 5, 7]$ and ${\bf{D}}=[1, 4, 6, 8]$, therefore MSB convolutional kernels would have receptive field of $[3\times3, 9\times9, 25\times25, 49\times49]$ over their input feature maps. Whereas, in the final level of the encoding path where feature maps are down-sampled by a factor of 8 and are in their smallest state, MSDU kernel sizes and dilation rates are to ${\bf{K}}=[3, 5, 3, 3]$ and ${\bf{D}}=[1, 1, 4, 6]$ resulting in receptive field sizes of $[3\times3, 5\times5, 9\times9, 13\times13]$ for MSB. This means that the convolutional kernels in our proposed MSDUs can extract relevant features starting from the scale of local structures size to the scale of nucleus size. We set the parameters of MSDUs heuristically based on the nuclei diameter analysis on the available data set.

\subsection{Proposal-Free Instance Segmentation}
\subsubsection{Segmentation and Centroid Detection}\label{seg-det}

For predicting mask and position of each nucleus, a dual-head network with similar architecture to SpaNet is utilized (see supplementary materials Fig.1). One head predicts the mask of nuclei, and another head predicts the centroids. The ground truth for predicting the centroids is built by considering each nucleus as a Gaussian-Shaped function where the maximum of Gaussian occurs at the center of the nucleus. The function \cite{koohababni2018nuclei} for constructing GT for each nucleus centroid on images, ${\bf{G}}_n$, is: 

\begin{equation}
  {{\bf{G}}_n {(x,y)}} = \left\{ \begin{array}{l}
\frac{1}{{1 + \beta \left\| {({c_{nx}},{c_{ny}}) - (x,y)} \right\|}}\,\,\,\,\,\,\,\,{\rm{if}}\,\,\,\,\,\left\| {({c_{nx}},{c_{ny}}) - (x,y)} \right\| \le r\\
0\,\,\,\,\,\,\,\,\,\,\,\,\,\,\,\,\,\,\,\,\,\,\,\,\,\,\,\,\,\,\,\,\,\,\,\,\,\,\,\,\,\,\,\,\,\,\,\,\,\,\,\,\,\,{\rm{elsewhere}},
\end{array} \right.
\end{equation}

where $({c_{nx}},{c_{ny}})$ and $(x,y)$ are the coordinate of  nuclei centroid and all possible coordinates of image pixels, respectively. 
In our experimentation $\beta$ and $r$ are, 0.01 and 8 respectively. Input to this network is RGB image, and we used smooth Jaccard and mean squared loss functions to minimize error for predicting mask and detection map respectively.

\subsubsection{Instance Segmentation}\label{sec:Instance}
An important part of instance segmentation is providing a GT that can reflect the separation between nuclei. 
To this end, we propose to use a GT tensor, ${\bf{P}}_{h\times w \times 6}$, that encompasses spatial information of all nuclei in the image. In ${\bf{P}}$, all pixels related to the $n$\textsuperscript{th} nucleus, are assigned with the same feature vector of spatial information, ${p_n}$.
 This vector is in the form of \cite{liang2017proposal}: ${p_n} = ({{{c_{nx}}} \mathord{\left/
 {\vphantom {{{c_{nx}}} w}} \right.
 \kern-\nulldelimiterspace} w},{{{c_{ny}}} \mathord{\left/
 {\vphantom {{{c_{ny}}} h}} \right.
 \kern-\nulldelimiterspace} h},{{{l_{nx}}} \mathord{\left/
 {\vphantom {{{l_{nx}}} w}} \right.
 \kern-\nulldelimiterspace} w},{{{l_{ny}}} \mathord{\left/
 {\vphantom {{{l_{ny}}} h}} \right.
 \kern-\nulldelimiterspace} h},{{{r_{nx}}} \mathord{\left/
 {\vphantom {{{r_{nx}}} w}} \right.
 \kern-\nulldelimiterspace} w},{{{r_{ny}}} \mathord{\left/
 {\vphantom {{{r_{ny}}} h}} \right.
 \kern-\nulldelimiterspace} h})$, where $({c_{nx}},{c_{ny}})$, $({l_{nx}},{l_{ny}})$, and $({r_{nx}},{r_{ny}})$ are the coordinates of the center, left top, and bottom right of the $n$\textsuperscript{th} nucleus'  bounding box, respectively. All the values are normalized by the width and height of bounding box, $(w,h)$. 
 A smoothed ${L_1}$ objective function that also ignores the background region in loss computation has been incorporated for the network optimization \cite{liang2017proposal}.
It is expected that the network predicts similar values for pixels belonging to the same nucleus. Note that the input to SpaNet for predicting nuclei spatial information has nine channels. The first six are made by concatenating RGB and HSV color channels, since nuclei are sometimes more distinguishable in HSV color space.
The remaining 3 channels are, predicted segmentation map (achieved in the previous step), ${\bf{M}}_{\rm{seg}}$, and spatial coordinate maps of pixels, $({\bf{M}}_{\rm{x}},{\bf{M}}_{\rm{y}})$. These last three channels inject the positional information to the SpaNet.

 \subsubsection{Post-Processing}
After predicting the spatial information of nuclei instances via SpaNet, we cluster them to attain the final instance segmentation. 
Directly clustering the predicted maps might fail due to the large spatial domain (number of pixels) and a high number of nuclei (number of clusters) in them. Therefore, we propose to apply the clustering algorithm on nuclei clumps separately. To identify these clumps, we firstly use a threshold the segmentation maps (section \ref{seg-det})   with a value of 0.3 and remove objects with an area smaller than 5 pixels to generate the nuclei masks. Connected components (CC) in the generated mask indicate isolated nuclei or nuclei clumps. By estimating the number of candidate nuclei (clusters) in a CC, we can start the clustering procedure.  The number of clusters per CC is determined by counting the number of local maxima in the intersection of that CC with the predicted detection map (section \ref{seg-det}). 
Similar to \cite{liang2018proposal} we use spectral clustering algorithm for it's effectiveness compared to other models by selecting Radial Basis Function kernel (RBF) as the affinity function.
 \section{Results and Discussion}
 \label{Sec:Results}
 \subsubsection{Dataset}
 The dataset consists of 30 images (16 for training and 14 for test set) from seven different tissues. Images were obtained from The Cancer Genome Atlas (TCGA) where 1000$\times$1000 patches were extracted from Whole Slide Images (WSIs) \cite{kumar2017dataset}. These seven tissues are kidney, stomach, liver, bladder, colorectal, prostate, and liver. Out of 14 test images, eight belongs to the same tissue type as the training set (seen organs), and six images are from different tissue types (unseen organs).
 For more details regarding dataset and test/train split refer to \cite{kumar2017dataset}.

 \subsubsection{Networks Setup}
  To attain generalization and robust predictions, we followed stochastic weight averaging approach proposed in \cite{izmailov2018averaging}. Cycling learning rate ($\alpha_i$) is adopted at each iteration $i$ as follows: ${\alpha _i} = (1 - {t_i}){\alpha _1} + {t_i}{\alpha _2}$,
  where ${t_i} = ({\rm{mod}}(i - 1,c) + 1) / {c}$, initial learning rate and final learning rate for each cycle are set to ${\alpha _1=0.01}$ and ${\alpha _2=0.0001}$, respectively, and cycling length is $c=20$ epochs.
Overall the network is trained for 100 epochs and the average of weights at the end of all cycles are computed for test time prediction.

All networks in the proposed framework have been trained using the same strategy, and "SGD" has been used as an optimizer to minimize objective functions. The input patch size for all networks is 256$\times$256. Networks for segmentation-detection and instance predictions are trained with a batch size of 2 and 4, respectively. Various data augmentation techniques were employed during the training of the network, their related details can be found in \cite{jahanifar2018segmentation}.

\subsubsection{Results and comparative analysis}
\begin{figure}[!t]
\centering
\includegraphics[width=\textwidth]{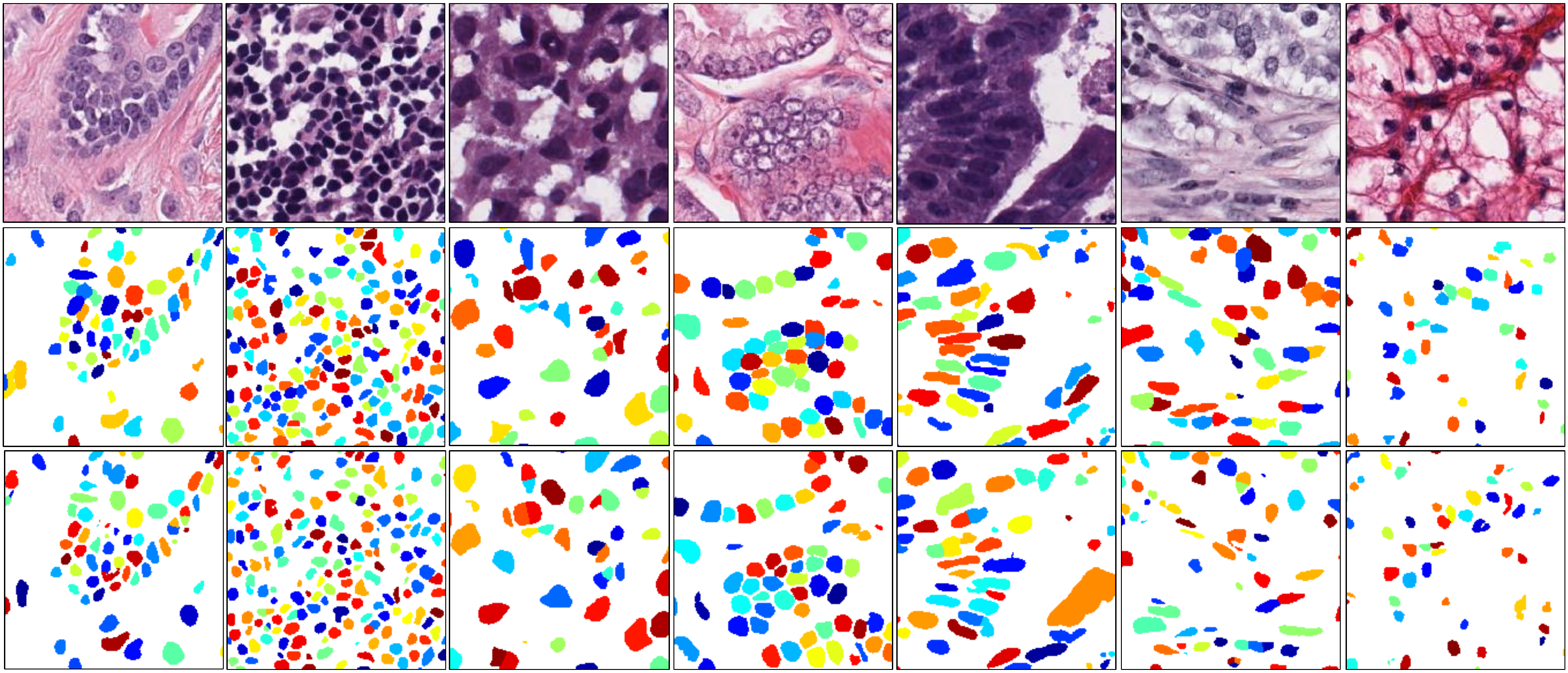}
\caption{Cropped images of seven different organs with their corresponding ground truth (second row) and prediction of our proposed method (third row).
}%
\label{fig:example}%
\end{figure}
\begin{table}[t]
\caption{Results of different methods on the nuclei instance segmentation test sets.}
\small
\begin{tabular}{c|c|c||c|c}
\hline
\multirow{2}{*}{Method} & \multicolumn{2}{c||}{AJI (\%)}                                                 & \multicolumn{2}{c}{F1-score (\%)}                                           \\ \cline{2-5} 
                        & \multicolumn{1}{l|}{Seen Organ} & \multicolumn{1}{l||}{Unseen Organ} & \multicolumn{1}{l|}{Seen Organ} & \multicolumn{1}{l}{Unseen Organ} \\ \hline
CNN3 \cite{kumar2017dataset}                    & 51.54                            & 49.89                                 & 82.26                            & 83.22                                \\ \hline
DR \cite{naylor2018segmentation}                    & 55.91                            & 56.01                                 & -                            & -                               \\ \hline
DCAN\cite{chen2017dcan}                & 60.82                           & 54.49                                & 82.65                           & 82.14                               \\ \hline
PA-Net \cite{liu2018path}                & 60.11                           & 56.08                                & 81.56                           & 83.36                               \\ \hline
Mask-RCNN  \cite{he2017mask}             & 59.78                          & 55.31                             &   81.07                               &  82.91                                    \\ \hline
BES-Net \cite{oda2018besnet}                 & 59.06                           & 58.23                                & 81.18                           & 79.52                               \\ \hline


CIA-Net  \cite{zhou2019cia}               & 61.29                           & 63.06                                & 82.44                           & \textbf{84.58}                               \\ \hline
Spa-Net (ours)                 &   \textbf{62.39}                        & \textbf{63.40}                                & \textbf{82.81}                          & 84.51                              \\ \hline

\end{tabular}
\label{tResults}
\end{table}

Performance of the proposed model is compared against several deep learning based methods as reported in Table \ref{tResults}. Except the baseline method (CNN3) \cite{kumar2017dataset} which categories the image pixels into three classes using a CNN-based classifier, other methods in Table \ref{tResults} (DR-Net \cite{naylor2018segmentation}, DCAN \cite{chen2017dcan}, BES-Net \cite{oda2018besnet}, and CIA-Net \cite{zhou2019cia}) took a dense prediction approach and used encoder-decoder like CNN.

As deduced from the results in Table \ref{tResults}, our proposed method based on SpaNet outperforms other state-of-the-art methods. Achieving AJI of 62.39\% and F1-score of 82.81\%  
shows an improvement of 1.10\% for AJI and 0.37\% for F1-score metrics compared to the best performing method in the literature. The superiority of the proposed method performance can be observed in both seen and unseen organs. Fig. \ref{fig:example} demonstrate the qualitative results of our method applied on all tissue types in test set.

The proposed framework offers several benefits. First, owing to the multi-scale and feature aggregation properties of MSDUs, using SpaNet architecture in this framework leads to more accurate instances' positional information. The performance of the current framework using off-the-shelf network architectures has also been shown in the supplementary material (see supplementary materials Fig. 5, Fig. 6). Second, in our proposed framework, we used separate models for predicting positional information and nuclei detection. Based on our experiments, considering a single network for all tasks (see supplementary materials Fig. 3) results in performance drop (see supplementary materials Fig. 4). Third, our proposed model incorporates much less number of parameters ($\sim$21M) in comparison with other models ($\sim$31M for U-Net and $\sim$40M for CIA-Net); therefore it has a better chance to generalize on unseen data (see supplementary materials Fig. 6). This is an important behavior in the current application with such a small data set.

\section{Conclusion}\label{Sec:Conclusion}
In this work, we presented a proposal-free framework for nuclear instance segmentation of histology images. Prediction of segmentation map, detection map, and spatial information of nuclei were aggregated in a principled manner to obtain final instance-level segmentation. To have a precise prediction, we proposed a spatial aware network which preserves the positional information throughout the network by incorporating a novel multi-scale dense unit. We showed that our method can achieve state-of-the-art performance on a multi-organ publicly available data set.

%
%
%
%

\bibliographystyle{splncs04}

\def\bibfont{\footnotesize}
\bibliography{MyCollection}
\end{document}